\documentclass[aps,nofootinbib,onecolumn,citeautoscript]{revtex4-2}
\usepackage{amsmath,amssymb}
\usepackage{enumerate}
\usepackage{algorithm}
\usepackage{algpseudocode}
\usepackage{xcolor}
\usepackage[colorlinks=true,citecolor=red,linkcolor=blue,linkbordercolor=green]{hyperref}
\usepackage[nameinlink,capitalise]{cleveref}
\usepackage{breqn}
\usepackage{setspace}
\usepackage{enumerate}
\usepackage{graphicx}

\begin{document}
\title{Tridiagonal Toeplitz Matrices and Bipartite Quantum Correlations}
\author{Varsha S. Sambhaje$^{1}$,
Suprabhat Sinha$^{2}$, Kapil K. Sharma$^{\star 3}$\\\vspace{0.4cm}
\textit{$^{(1,2,3)}$D Y Patil International University,\\ Akurdi, Pune, Maharashtra 411044, India}\\
E-mail: $^{1}$varsha.sambhaje@dypiu.ac.in, $^{2}$suprabhatsinha64@gmail.com,  $^{3}$anju.chaurasia@dypiu.ac.in,  $^{\star 4}$iitbkapil@gmail.com}

\begin{abstract}
In this article, we focus on tridiagonal Toeplitz Hermitian matrices, which fulfill the requirement of a valid Hamiltonian often used in Quantum Information. We investigate the behavior of such matrices to pursue the dynamics of quantum correlations (entanglement and quantum discord) for bipartite Werner state and maximally entangled mixed states. We have found interesting results that the main diagonal terms in the Toeplitz matrices never affect the quantum correlations in both quantum states. However, super-diagonal and sub-diagonal terms play the important role in the dynamics. We investigate the phenomenon of entanglement sudden death and also observe the presence of quantum discord in the absence of entanglement. Most importantly it is found that MEMS is more sensitive in comparison to the Werner state.
\end{abstract}


\maketitle

\section{Introduction}

Quantum correlations are the backbone of quantum computation and information; their dynamical study is central to this discipline \cite{nl}. To execute many quantum applications like quantum teleportation, super-dense coding, quantum image processing, quantum machine learning, and quantum algorithms \cite{QC,KKSM,QT1,QT2,QIMGP,qm}; the quantum correlations are primary requirement. There are several kinds of quantum correlations but here we are focusing on concurrence and quantum discord \cite{ESD,QD1}. To pursue the dynamical study of quantum correlations, we need mathematical tools for their detection and quantification \cite{EM, Ad}. We have enough mathematical tools to quantify and detect such quantum correlations in bipartite Hilbert space but in higher dimensions, (i.e $d\otimes d)$; the development of such tools is still a challenging mathematical problem. It is always an essential requirement to maintain the sustainability of quantum correlations in a physical system for a long time. However, these correlations are very much sensitive and entanglement often passes through sudden death as time evolves; this phenomenon is called entanglement sudden death (ESD) \cite{ESD1}. It has been observed that in the absence of entanglement, quantum discord has existence but there are limited studies to use quantum discord for quantum applications in a practical scenario. Calculating quantum discord is an NP-complete problem \cite{QD2}. A bipartite system is the primary one, which can be the subsystem of a larger system. In the context of quantum application development, if ESD takes place then no quantum application will produce the result. Therefore it is always interesting to investigate ESD in a physical system under certain conditions. ESD also has been observed in varieties of quantum states under Dzyaloshinsky-Moriya interactions \cite{KKS2, KKS3, KKS4, KKS5, scr}.

Toeplitz matrices have significant applications in classical as well as in quantum domains \cite{T}. Toeplitz operators in quantum mechanics are known for a long time in literature and their special properties have been studied in all kinds of Hilbert spaces such as Banach, Hardy, and Fock spaces, etc. \cite{TO, B, B1}. The theory of the Toeplitz determinant is developed in 1967, which is used in calculating the spin correlations in the two-dimensional Ising model \cite{TS1, TS2}. In 2017 the properties of these matrices have been studied in Long-Range Kitaev spin chains to investigate topological phase transition \cite{K}. Such metrics are around in quantum mechanics at a larger scale although in quantum chemistry \cite{TQC}. Often if the Hamiltonian of a physical system is a large dimensional sparse Hermitian matrix then such Hamiltonian is mapped to tridiagonal Toeplitz matrices by using the Lanczos algorithm to simplify the calculations \cite{L1, L2, L3}. The Lanczos algorithm is numerically stable if the elements of a given Hamiltonian are non-floating numbers \cite{NSL}. 

From a quantum algorithms perspective, recently these matrices have had a prominent role in the development of the quantum imaginary time evolution (QITE) model to do progress in quantum algorithms \cite{QITE1,QITE2,QITE3,QITE4}. In QITE we deal with a larger dimensional k-local many-body Hamiltonian i.e. $(H=\sum_{i}^{n}h_{i}\in \mathbb{B}((\mathbb{C}^{d})^{\otimes n})$, where each $h_{i}$ is k-local. This Hamiltonian is used to find out the quantum imaginary time evolution by assuming time as an imaginary component \cite{klocal}. The time evolution operator is given as $e^{iHt}$ which is non-unitary. Further, this non-unitary operator is approximated by choosing the appropriate k-Local Hamiltonian $H$ of the system and mapped it to a unitary operator by using Trotter expansion with errors \cite{QITE1}. The error is reduced by considering the nearest neighbour interactions by selecting the proper k-local Hamiltonian. This scheme provides a better convergence time of the expectation value of energy in comparison to the variational classical-quantum approach \cite{VQA}. However, selecting the better k-Local Hamilton with less amount of error is an art and there is no perfect method known for the same. In general, the Hamiltonian matrix considered for QITE has a very large dimension, so in this case, one need to map the given matrix onto tridiagonal Toeplitz form by using recently proposed Quantum Lanczos algorithms \cite{QITE1}, which maps a k-local Hamiltonian to Toeplitz matrices.  

Recently Farenick et al. developed the Toeplitz operator structure over the kronecker product of bipartite Hilbert space by using the $C^{*}$ algebra; further, they have studied entanglement in Toeplitz matrices \cite{F1, F2}. Following the literature on Toeplitz matrices, it is very important to investigate the dynamical behavior of quantum correlations under such matrices. Here we mention that the impact of tridiagonal Toeplitz matrices on different quantum states is missing in quantum information. Hence we plan to investigate the role of tridiagonal Toeplitz (special case) matrices to study the dynamics of quantum correlations in bipartite Werner state and maximally entangled mixed states (MEMS) \cite{WS, MEMS}. In the present study we consider the specific form of such matrices with the elements as exponential functions of the parameters i.e. $(a^{n},b^{n})$, the reason to choose such elements is that by putting the parameter values $(a,b,n)$, different Hamiltonian can be produced with numerical entries, hence it generalize the results. 

The paper is organized as follows, in section II, we present the tridiagonal Toeplitz Hamiltonian with its eigenvalues and eigenvectors. Also, we discussed the unitary time evolution density matrix. Section III gives information on initial quantum states. In section IV, we provide the quantum correlation measures. Further, section V is dealing with the eigenvalue spectrum and density matrices for quantum dynamics. In section VI, we discuss the results based on the dynamical behavior of the Werner state and MEMS respectively. In the last section VII, we provide the conclusion. 

\section{Toeplitz matrices and unitary time evolution}
In this section, we present the Hamiltonian of the physical system, which has the structure of a tridiagonal Toeplitz matrix with ($H=H^{\dagger}$). This matrix has three diagonals namely main diagonal, super-diagonal, and sub-diagonal. The reader can go through the formation of the tridiagonal Toeplitz operator with the references therein \cite{F2}. The structure of the matrix considered  for the present study is given below, 

\begin{equation}
H_{n}=
\begin{pmatrix}
a^{n} & b^{n} & 0 & 0 \\
b^{n} & a^{n} & b^{n} & 0 \\
0 & b^{n} & a^{n} & b^{n} \\
0 & 0 & b^{n} & a^{n}
\end{pmatrix}\label{H},
\end{equation}
where $(a,b,n)\in R$. The eigenvalues and eigenvectors of $H$ are given below, \\
\begin{equation}
  \lambda_{j} = b^{n} + 2a^{n} \sqrt{\frac{b^{n}}{a^{n}}}\cos{\frac{j\pi}{m + 1}}, \quad \quad with \quad (j, m) = (1, 2, 3, 4)
\end{equation}
and 
\begin{equation}
x_{j}=  \left(\frac{ b^{n}}{a^{n}}\right)^{\frac{m}{2}} \sin \left( \frac{{mj\pi}}{m+1}\right), \quad \quad with \quad (j, m) = (1, 2, 3, 4)
\end{equation}

where $\lambda_{j}$ are the eigenvalues and $x_{j}$ are the corresponding eigenvectors.

In general, the entries of a Hamiltonian give information about the interaction strength in a physical system. We are intended to look at how fast the dynamics of quantum correlations change as the interaction strength in $H_{n}$ grows exponentially; hence we recall the reason that we choose the entries of the $H_{n}$ as $(a^{n},b^{n})$.

According to the postulates of quantum mechanics \cite{scr}, the time-dependent Schrödinger equation shown below governs the unitary dynamics of the physical system,

\begin{equation}
i\hbar\frac{d}{dt}\vert \psi(t)\rangle=H \vert\psi(t)\rangle,
\end{equation}
where $H$ is the Hamiltonian of the physical system with ($H=H^{\dagger}$). The solution of this equation is obtained as,
\begin{equation}
\vert\psi(t)\rangle=e^{\frac{-iHt}{\hbar}}\vert\psi(0)\rangle.
\end{equation}
Using the above equation, we can obtain the time evolution density matrix as follows,
\begin{equation}
\rho(t)=U(t).\rho(0).U(t)^{\dagger},\label{tm1}
\end{equation}
where $U(t)=e^{\frac{-iHt}{\hbar}}$ is the unitary matrix that has the Hamiltonian $H$ in the exponential. Here we assume ($\hbar=1$) to maintain the simplicity of the current study. The Eq.$(\ref{tm1})$ is used in current work to develop the dynamics of quantum correlations in Werner state and MEMS.

We prepare the initial state of the bipartite system in Werner state and MEMS respectively. The density matrices of these quantum states are formed by the two qubits provided in the next section. 

\section{Initial quantum states}
In this section, we present the mathematical structure for the Werner state \cite{WS} and MEMS \cite{MEMS} with their corresponding density matrices. Both states are parametrized; hence their quantum correlations are the functions of parameters involved in these states. The respective quantum states are described in successive subsections.

\subsection{Werner state}
The Werner state is invariant under unitary transformation in $d-$dimensional Hilbert space. This property reads as below, 


\begin{equation}
\rho=(U\otimes U)\rho(U^{\dagger}\otimes U^{\dagger}),
\end{equation}
where $\rho$ denotes the density matrix of the system. The Werner state adopt the following form while dealing with two qubits, 
\begin{equation}
\rho^{WS}=\gamma\vert\psi^{-}\rangle\langle\psi^{-}\vert+(1-\gamma)\frac{I}{4} \quad with \quad (0\leq \gamma \leq 1)\label{w1},
\end{equation}
where $\vert\psi^{-}\rangle$ indicates a singlet state and I is the ($4 \times 4$) dimensional identity matrix. The matrix form is expanded below,
\begin{equation}
\rho^{WS}=\left(
\begin{array}{cccc}
 \frac{1-\gamma }{4} & 0 & 0 & 0 \\
 0 & \frac{\gamma +1}{4} & -\frac{\gamma }{2} & 0 \\
 0 & -\frac{\gamma }{2} & \frac{\gamma +1}{4} & 0 \\
 0 & 0 & 0 & \frac{1-\gamma }{4} \\
\end{array}
\right)
\end{equation}


\subsection{Maximally entangled mixed states (MEMS)}
Another bipartite quantum state that is closely related to the bipartite Werner state is MEMS. The two qubits MEMS, which is empirically proven as well as being more entangled than the two qubits Werner state in terms of concurrence measure \cite{C1, C2}, the state is prepared by Munro et al. \cite{MEMS}. Two-qubits bipartite MEMS' density matrix can be expressed as,
\begin{equation}
\rho^{MEMS}=\left[\begin{array}{cccc}
g(\gamma) & 0 & 0 & \frac{\gamma}{2}\\
0 & 1-2g(\gamma) & 0 & 0\\
0 & 0 & 0 & 0\\
\frac{\gamma}{2} & 0 & 0 & g(\gamma)
\end{array}\right]\label{m1}.
\end{equation}
The above state incorporates a function $g(\gamma)$ with following conditions,
\begin{equation}
g(\gamma)=\delta=\left\{\begin{array}{cc}
\frac{1}{3}, &\quad 0\leq\gamma<\frac{2}{3}\\
\frac{\gamma}{2}, &\quad \frac{2}{3}\leq\gamma\leq1
\end{array}\right .\nonumber.
\end{equation}
For the sake of convenience, we will refer to $g(\gamma)$ as $\delta$ throughout this study. 

To proceed with the dynamical study in specified quantum states, we need theoretical measures to quantify the entanglement and quantum discord. These measures are given in the next section.

\section{Measures of quantum correlations}
The quantum correlation will provide information about the entanglement of the system. In this section we explore the quantum correlation of the system with the help of two important measures of quantum correlations viz., concurrence and quantum discord \cite{QD1}. 

\subsection{Concurrence}
Concurrence is one of the quantum correlation measure that is widely used to quantify the entanglement of bipartite quantum systems \cite{C1, C2}. Concurrence of a bipartite quantum system governed by the corresponding density matrix $\rho$ can be defined as,
\begin{equation}
C(\rho)=max\,\{0,\lambda_{1}-\lambda_{2}-\lambda_{3}-\lambda_{4}\}\label{con},
\end{equation}
where $\lambda_{i}$'s are the decreasing order square root of eigenvalues of $\rho\tilde{\rho}$ and
\begin{equation}
\tilde{\rho}=(\sigma_{y}\otimes\sigma_{y})\rho^{*}(\sigma_{y}\otimes\sigma_{y}),
\end{equation}
where $\tilde{\rho}$ is the result of the spin-flip operation applied on density matrix $\rho$. Here $\sigma_{y}$ is the Pauli Y matrix and $\rho^{*}$ is the complex conjugate of the density matrix $\rho$.
\subsection{Quantum discord}
Quantum discord is another measure of quantum correlations which characterized mathematically in terms of quantum mutual information. In the classical domain the mutual information between two random variables $[X, Y]$ can be expressed in  equations given below,
\begin{eqnarray}
I(X;Y)=H(X)+H(Y)-H(X,Y), \\
I(X;Y)=H(X)-H(X \vert Y).
\end{eqnarray}
Here we mention that both the above equations are different in the quantum domain as they depends on quantum measurements. Considering a bipartite composite density matrix $\rho$, incorporating $\rho_{A}$ and $\rho_{B}$, we can express the quantum version of the above equations as follows,
\begin{eqnarray}
I(\rho)=S(\rho_{A})+S(\rho_{B})-S(\rho)  \label{qd1} \\
I(\rho)=S(\rho_{A})-S(\rho_{A}\vert \rho_{B}) \label{qd2}
\end{eqnarray}
The term $S(\rho_{A} \vert \rho_{B})$ is obtained by performing the quantum measurement on either sub-system $A$ or $B$ of the composite system $AB$. It is important to note that quantum discord is not symmetric with respect to the measurement performed. Here we consider that the measurement $M$ is performed on subsystem $A$, so in this direction the term $I(\rho \vert M)$ is obtained as,
\begin{equation}
I(\rho\vert M)=S(\rho_{B})-S(\rho\vert M)).
\end{equation}
We have to maximize the term $I(\rho \vert M)$ over all possible measurements, and the alternate form of the expression $I(\rho \vert M)$ can be given by,
\begin{equation}
C(\rho)=S(\rho_{B})-\text{min}_{M}S(\rho\vert M).
\end{equation}
The difference between the expressions $I(\rho)$ and $C(\rho)$ is called quantum discord, which is given below,
\begin{equation}
Q(\rho)=I(\rho)-C(\rho)=S(\rho_{A})-S(\rho)+\text{min}_{M}S(\rho \vert M).
\end{equation}

\section{Eigenvalue spectrum and density matrix}
We proceed to develop the dynamical form Eq.($\ref{tm1}$) by using the Hamiltonian given in Eq.(\ref{H}). This dynamical equation is explored with ($\rho(0) = \rho^{WS}$) and ($\rho(0) = \rho^{MEMS}$) respectively. 
For Werner state the eigenvalue spectrum of $\rho^{W}(t)$ is obtained as,
\begin{equation}
\left\{\frac{1-\gamma }{4},\frac{1-\gamma}{4},\frac{1-\gamma }{4},\frac{1}{4} (3\gamma +1)\right\}\label{W}
\end{equation}
We observe that the above eigenvalue spectrum is free from all the Hamiltonian parameters $(a,b,n,t)$, hence the concurrence and quantum discord are never affected in Werner state from these parameters.
 
 We have found that the eigenvalue spectrum $\rho(t)$ of MEMS is very complicated to calculate analytically as it is a function of the parameters $(a, b, n, \gamma, t)$. Further, we pursue the numerical study and presented the density matrix $\rho^{MEMS}(t)$ as below.
\begin{equation}
\left(
\begin{array}{cccc}
 \frac{a_1+b_1}{5}  &  \frac{\phi_{6}P_1}{20}  &  e_{7}[n_3 + n_4 + n_5]  &  \frac{c+d}{10} \\ \\
 e_6 [m_1 \theta_1 + n_1 \theta_2 + m_3 \phi_4]  &  \frac{-d_1 [e_8 P_3 + P_4]}{10}  &  e_8 [P_5 + P_6 + P_7]  &  e_6 m_5 \\ \\
e_6 [n_1 \theta_1 -m_1 \theta_2 + n_2 \phi_4]   &  e_8 [P_6 -P_5 -P_7]   &  e_8 [P_5 - m_4 + P_8]   &  e_6 m_5 \\ \\
 \frac{c - d}{10} & \frac{\phi_6 P_2}{5} & e_7 [n_3 -n_4 + n_6] & \frac{a_1 - b_1}{5} \\ \\
\end{array}
\right)
\end{equation}
where, the terms of matrix are defined as follows,\\
\begin{align*}
a_1 = 1 + \delta + d_2 \phi_3 , \quad b_1 = d_1 \phi_1 (-1 + \phi3) , \quad  c = \gamma (3 + 2 \phi_3) ,\quad d = 2 i d_1 \phi_2 (-1 + \phi_3),
\end{align*}
\begin{align*}
P_1 = -(-1+ \sqrt{5}) e^{\frac{ie_3}{2}} (\gamma -2 \delta) \phi_5 +  e^{\frac{-ie_3}{2}}[-(-1+ \sqrt{5} +e_1 + \sqrt{5} e_1 ) \gamma \phi_5 + 2(1 - \sqrt{5} + e_2 + \sqrt{5} e_2) \delta \phi_5] \\ - 4 \phi_4 (\gamma \phi_4 + 2i (-1 + \delta) \phi_5) \phi_6 + 4 \sqrt{5} \phi_4 (2i (-1 + 3 \delta) \phi_4 + \gamma \phi_5)\phi_7
\end{align*}
\begin{align*}
P_2&=(-d_1 + d_1 \phi_1) \phi_6 + \sqrt{5} (i \gamma + d_1 \phi_2 )\phi_7,           &  P_3&=[h_1 e_3 + h_1 - h_2 e_1 ]^{2},\\
P_4&=\frac{\delta}{5}[ 2 - e^{-i \sqrt{5}b^{n}t} + e^{i \sqrt{5}b^{n}t}],    &  P_5&=d_1 h_3 -d_1 e^{2i e_3}h_3 + d_1 e^{2i b^{n}h_4},\\
P_6&=d_1 e^{2i \sqrt{5}b^{n}t}h_4 + 8 e^{i e_3}\gamma -4 e^{ie_4}\gamma -4 e^{ib^{n}t}\gamma, & P_7&=4d_1 e^{i \sqrt{5}b^{n}t} - 4d_1 e^{ie_5},\\
P_8&=4 d_1 e^{ie_5} + 4d_1 e^{i \sqrt{5}b^{n}t}, & e_1&=e^{i b^{n}t} + e^{i \sqrt{5}b^{n}t},\\  e_2&=e^{i b^{n}t} - e^{ \sqrt{5} b^{n }t}, &
e_3&=(1 + \sqrt{5}) b^{n}t,\\
e_4&=(1 + 2 \sqrt{5}) b^{n}t, & e_5&=(2 + \sqrt{5}) b^{n}t,\\
e_6&=\frac{1}{20 \sqrt{5}} e^{\frac{-i e_3}{2}}, & e_7&=\frac{1}{20 \sqrt{5}} e^{\frac{-i e_5}{2}} \phi_6,\\ e_8&=\frac{1}{40} e^{-i e_3}, &  d_1&=-1 + 2 \delta, \\ d_2&=-1 + 4 \delta, & h_1&=5 + \sqrt{5},\\ 
h_2&=-5 + \sqrt{5}, & h_3&= 3 + \sqrt{5},\\  h_4&=-3 + \sqrt{5}, & \theta_1&=2 \delta cos(\frac{b^{n} t}{2}) - i \gamma sin(\frac{b^{n} t}{2}),\\
\theta_2&=\gamma cos( \frac{b^{n} t}{2}) - 2 \delta i  sin(\frac{b^{n} t}{2}), & \theta_3&= i \gamma cos(\frac{b^{n} t}{2}) + 2 \delta  sin(\frac{b^{n} t}{2}),\\ \theta_4&=2i \delta cos(\frac{b^{n}t}{2}) + sin(\frac{b^{n}t}{2}), &  \phi_1&=cos(b^{n} t),\\ \phi_2&= sin(b^{n} t), & \phi_3&=cos(\sqrt{5}b^{n}t),\\ \phi_4&=cos(\frac{b^{n} t}{2}), & \phi_5&= sin(\frac{b^{n} t}{2}),\\ \phi_6&=sin(\frac{1}{2}\sqrt{5}b^{n} t), & \phi_7&=cos(\frac{1}{2}\sqrt{5}b^{n} t),\\  m_1&=(-5 + \sqrt{5}) e^{i (1 + \sqrt{5}) b^{ n} t} - (5 + \sqrt{5}) (e^{i b^{ n} t} +  e^{i  \sqrt{5} b^{ n} t}) + h2), & m_2&=(-5 + \sqrt{5}) + e_2 h_1 - e^{ie_3} h_2,\\ m_3&= 2[h_1 e^{ie_3}- h_2 e_1 + h_1]d_1, & m_5&=[m_1 \theta_3 - m_2 \theta_4 + m_3 \phi_5],\\ m_4&= d_1 e^{2i \sqrt{5}b^{n}t}h_4 + e^{ie_4}(4 - 16 \delta) + e^{ib^{n}t}(4 - 16 \delta) + e^{ie_4}(12 - 8 \delta),\\
n_1&=-(h_2 + e_{2}h_1 - e^{ie_3}h_2), & n_2&=2(-h_1e^{ie_5} + h_2 e_2 + h_1),\\ n_3&=-d_1 h_1 + d_1 e^{i \sqrt{5}b^{n}t}h_2, & n_4&=d_1 e^{ie_5} h_1 -  d_1 e^{2i b^{n}t}h_2,\\ n_5&=2 e^{ie_3}[ \sqrt{5} d_2 + 5 \gamma] +2 e^{i b^{n}t}[5 \gamma - \sqrt{5} d_2], & n_6&=2 e^{ie_3}[ \sqrt{5} \gamma + 5 d_2] +2 e^{i b^{n}t}[5 d_2 - \sqrt{5}\gamma]
\end{align*} 

By looking at the elements of density matrix, we have found that the matrix is free from the parameter $a$. Here the diagonal elements of tridiagonal matrix does not play the role in the dynamics. 

\section{Quantum dynamics for Werner state and MEMS}
Based on the results derived in the simulation of previous sections, here we present the dynamical aspects of quantum correlations (Concurrence and quantum discord). We recall that the eigenvalue spectrum of Werner state i.e. $\rho^{W}(t)$ given in Eq.(\ref{W}) is free from all the parameters $(a,b,n, t)$ hence the quantum correlations in Werner state are unaffected under the tridiagonal Toeplitz matrix.

Further, we present the density matrix $\rho^{MEMS}(t)$ of MEMS for the value $(t=0)$, the $\rho^{MEMS}(t)$ maps to $\rho^{MEMS}(0)$ and eigenvalue spectrum is obtained as below for $(\delta=\frac{1}{3})$ and $(\delta=\frac{\gamma}{2})$ respectively,
\begin{equation}
\left\{\frac{1}{3},0,\frac{1}{6} (2-3 \gamma ),\frac{1}{6} (3 \gamma +2)\right\},
\end{equation}
\begin{equation}
\{0,0,1-\gamma ,\gamma \}
\end{equation}

\begin{figure*}
\centering
\includegraphics[scale=0.84]{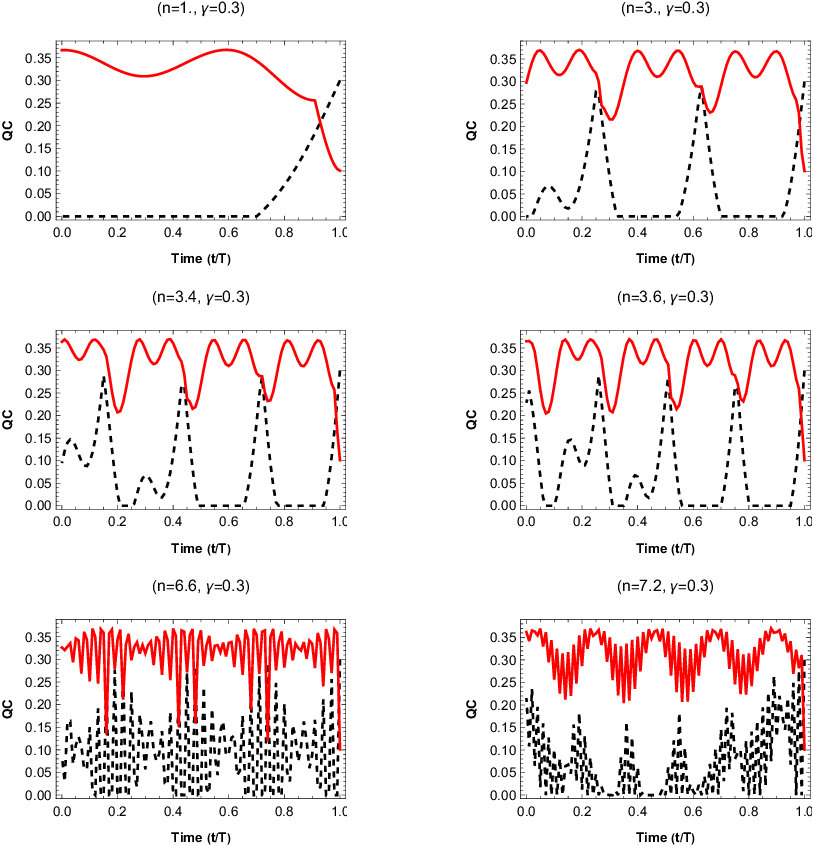} 
\caption{Plot of quantum corrections for (b=2) and $(\gamma$=0.3)}\label{f_1}
\end{figure*}

Here, we observe that both the eigenvalue spectrums are free from all the parameters $(n, a,b,t)$, and hence the initial quantum correlations are never affected by these parameters; these remain the same as in $\rho^{MEMS}(0)$. It is observed in this state that the quantum correlations are very sensitive with the parameters $(b,n,\gamma)$, and independent of the parameter $a$. Further we divide the study into two subsections with different parameter ranges. 

\begin{figure*}
\centering
\includegraphics[scale=0.84]{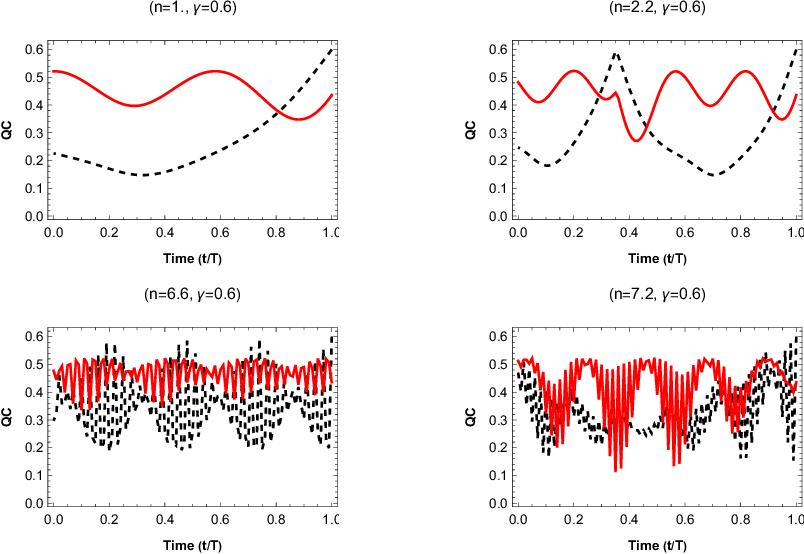} 
\caption{Plot of quantum corrections for (b=2) and $(\gamma$=0.6)}\label{f_2}
\end{figure*}

\subsection{Entanglement and quantum discord in MEMS with $(b=2)$ and varying $(n,\gamma)$.}
\begin{figure*}
\centering
\includegraphics[scale=0.84]{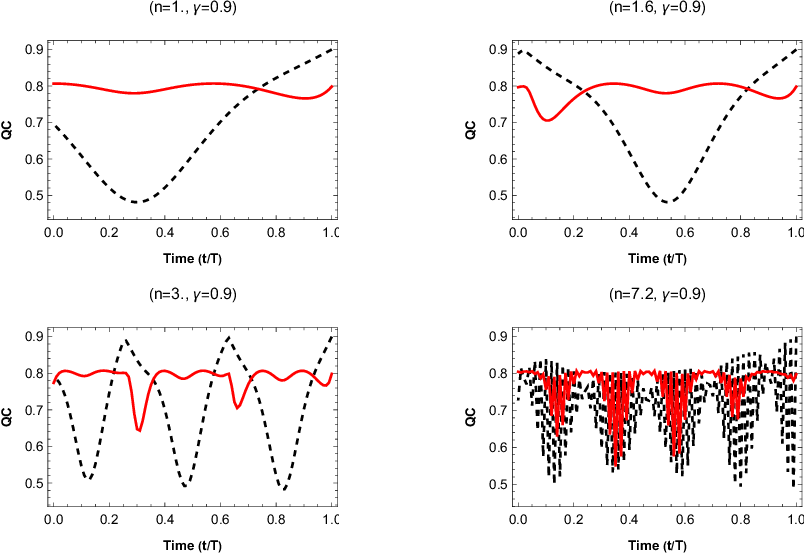} 
\caption{Plot of quantum corrections for (b=2) and $(\gamma$=0.9)}\label{f_3}
\end{figure*}

In this subsection, we present the results with the fixed value of $b$ and vary the parameters $(n, \gamma)$.  These results are presented in \cref{f_1,f_2,f_3}. The concurrence is depicted by black dotted color and quantum discord is sketched by a red solid color.  

\subsubsection{Case 1: $(\gamma=0.3)$}
In \cref{f_1}, we plot the results for the parameter $(b=2)$ with varying values of $n$ (keeping $\gamma$ fixed). With the specified values of parameters $b$ and $n$ the super-diagonal and sub-diagonal of the tridiagonal Toeplitz Hamiltonian becomes an exponential function $2^{n}$ which rises exponentially w.r.t to $n$.  We investigate ESD with the initial values of $(n=1,\gamma=0.3)$. Further when the value of $n$ increases the periodic ESD takes place but the width of the ESD zone decreases. In the absence of entanglement quantum discord exists that can be used for quantum applications. We also observe the increasing value of $n$ increases the frequency of quantum correlations (entanglement, quantum discord). For ($n=7.2, \gamma=0.3$), the frequency of oscillations is very high and quantum discord achieves a higher amplitude than entanglement.

\subsubsection{Case 2: $(\gamma=0.6)$}
In \cref{f_2},  we obtain the results for $(\gamma=0.6)$ and varying $n$. Following the previous case, it is observed here in the current case that the increasing values of $\gamma$ make disappear the sudden death of entanglement and rise the amplitude of quantum correlations. The initial amplitude of quantum discord at $(t=0)$ is approximately $0.5$ with $(n=1)$. Here we mention that increasing values of $n$ freeze the initial amplitude of quantum discord but this amplitude of entanglement rise with increasing values of $n$. For $(n=7.2)$, the entanglement achieves the initial amplitude as $0.48$ which is still less than the previously mentioned initial amplitude of quantum discord. It is interesting to note that the increasing values of $n$ increase the oscillations of both the quantum correlations.

\begin{figure*}
\centering
\includegraphics[scale=0.84]{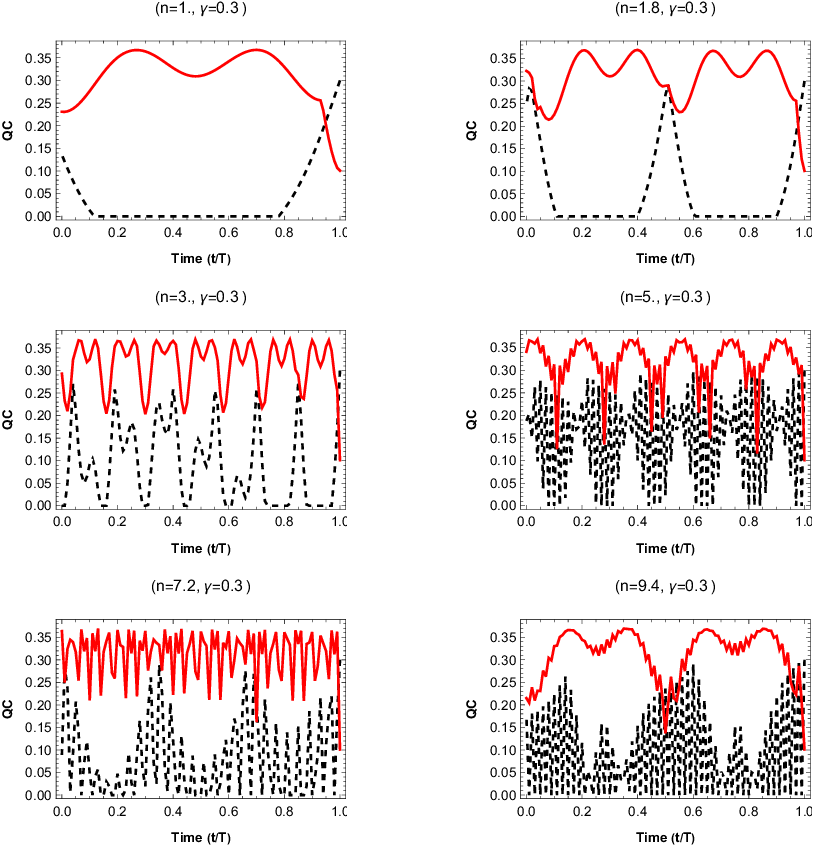} 
\caption{Plot of quantum corrections for (b=e) and $(\gamma$=0.3)}\label{f_4}
\end{figure*}

\subsubsection{Case 3: $(\gamma=0.9)$}
In \cref{f_3}, we plot the graphical results for $(\gamma=0.9)$. We revise that the elements of MEMS are the functions of parameter $\gamma$, so the increasing value of this parameter generates different quantum states (MEMS). With the value $(n=1,\gamma=0.9)$, the initial amplitude of quantum discord is approximately $0.8$ and for entanglement, it is $0.7$. We also observed that the initial amplitude of quantum discord freeze with varying $\gamma$. Next, we observed an interesting behavior that at $((t/T)=0.7)$, the amplitude of entanglement overshoots the quantum discord as time advances, the crossing of quantum correlations are phrased as a balancing point \cite{scr}.  As the value of $n$ increases the number of balancing points increases.

\subsection{Entanglement and quantum discord in MEMS with $(b=e)$ and varying $(n,\gamma)$.}
In this section, we present the results for $(b=e)$ and varying values of $(n,\gamma)$. The graphical results are depicted in \cref{f_4,f_5,f_6}. When we consider $(b=e)$, the elements of tridiagonal Toeplitz Hamiltonian take the form of the exponential function as $e^{n}$, which is the upper bound of an exponential function $2^{n}$ and have the higher growth rate. This study is taken over in subsequent cases with the varying value of $\gamma$.
 
\subsubsection{Case 1: $(\gamma=0.3)$}
In this subsection, we discuss the results in \cref{f_4}. With the parameter values $(n=1,\gamma=0.3)$ we obtain the sudden death results with the time interval $(0.1\leq t\leq 0.78)$. Moreover, as the value of $n$ increases the frequency of ESD and quantum discord increases. Here we mention that at $(t=0)$ the amplitude of quantum discord is less as compared to case 1 of $(b=2^{n})$ also the amplitude of entanglement is higher and goes on decreasing with time. Further, we have found the two sudden death zones with the parameters $(n=1.8,\gamma=0.3)$.

\subsubsection{Case 2: $(\gamma=0.6)$ and $(\gamma=0.9)$}
\begin{figure}
\centering
\includegraphics[scale=0.84]{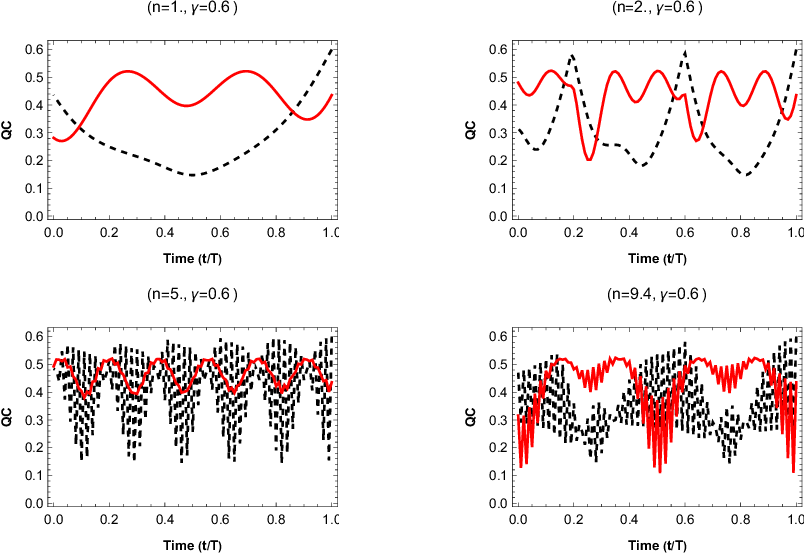} 
\caption{Plot of quantum corrections for (b=e) and $(\gamma$=0.6)}\label{f_5}
\end{figure}

\begin{figure}
\centering
\includegraphics[scale=0.84]{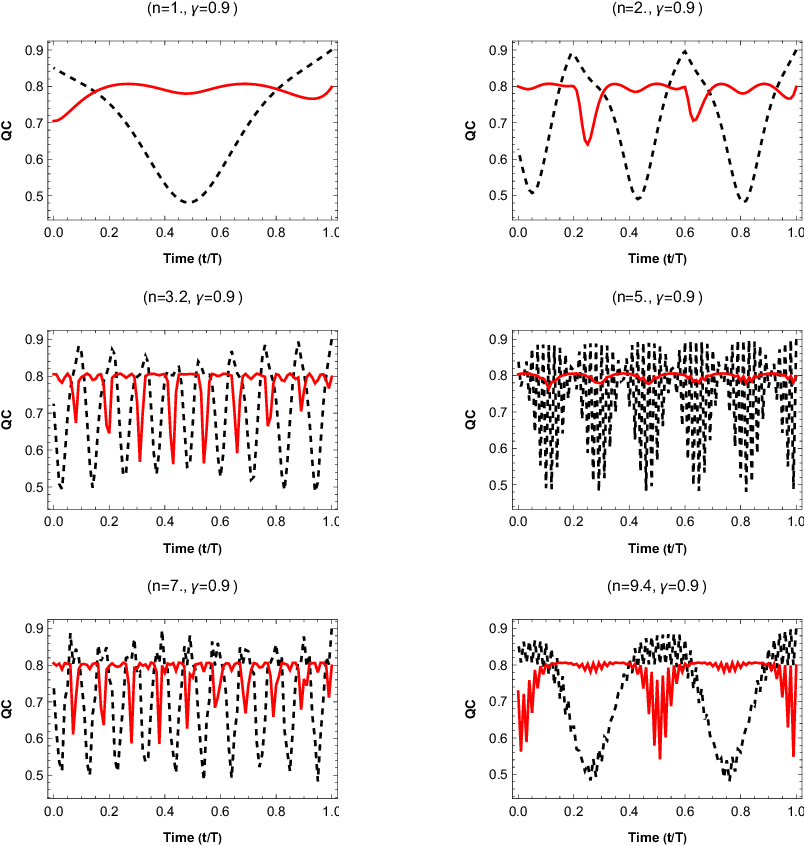} 
\caption{Plot of quantum corrections for (b=e) and $(\gamma$=0.9)}\label{f_6}
\end{figure}

In this subsection, we explore the dynamics with the parameters $(\gamma=0.6,\gamma=0.9)$ in \cref{f_5,f_6}. We have found that as the condition $(\gamma\geq 0.6)$ satisfied, the sudden death disappears in the system. Further, we mention that the frequency and amplitude behavior remains the same as discussed in previous subsections. 

For the value $(\gamma=0.9)$, there is a trade-off between the initial amplitude of entanglement and quantum discord. With the values $(n=1,\gamma=0.9)$, the initial amplitude of entanglement is higher than quantum discord, but with $(n\geq 2)$, the quantum discord at $(t=0)$, achieves higher amplitude than entanglement and this behavior remain same with the range $(2\leq n\leq 7)$. Further with $(n=9.4,\gamma=0.9)$, the initial behavior of entanglement dominates the quantum discord. As time advances the sinusoidal pattern of quantum correlations arises. 

\section{Conclusion}
In this work, we present the analysis of the quantum dynamics in two different kinds of quantum states i.e. Werner state and MEMS respectively. The study is focused to investigate the role of tridiagonal Toeplitz Hamiltonian in quantum information. We have considered the generalized elements of such Hamiltonian as the  exponential functions $(a^{n},b^{n})$ and found interesting results under unitary dynamics. We analysed that the eigenvalue spectrum of the Werner state is free from all the parameters $(a,b,n)$, hence the quantum correlations in this state are unaffected by the Hamiltonian parameters. Further for MEMS, the eigenvalue spectrum is independent of diagonal parameters $(a^{n})$, it is depending only on $(b^{n})$ and the parameter $\gamma$. We explored this study with the two different cases $(b=2)$ and $(b=e)$. We noticed that the quantum dynamics in MEMS are more fragile under such parameter changes and it passes through sudden death of entanglement. The frequency of quantum correlations in this state increases as the value of the parameter $n$ increases. Further we also investigated the balancing points of quantum correlations. The present study can be explored in more detail by expanding the range of the parameters $(a,b,n)$ in Toeplitz matrices, and hope it will be beneficial for the quantum information community.
\section*{Compliance with ethical standards}
The present work does not use any kind of experimental or human data. Also the work does not have any financial or competing interest.
\appendix
\section{Lanczos algorithm}
\begin{algorithm}[H]
\caption{Lanczos Algorithm}\label{alg:cap}
\begin{algorithmic}[1]
\Require Let $B \in F^{n \times n}$ be a Hermitian matrix. This algorithm computes the Lanczos iterations, i.e., an orthonormal basis $V_{m} = [v_{1}, . . . , v_{m}]$ for $V_{m}(x)$ where m is the smallest index such that $V_{m}(x) = V_{m+1}(x)$ , and the tridiagonal Toeplitz matrix $T_{m}$
\State $v_{1} \in C^{n}$
\State $q := \frac{x}{k \times k}; Q_{1} = [q]$;
\State $r := Aq$;
\For{$j = {2, 3, . . .}$}
\State $q := \frac{w_{1}}{ \beta_{j-1}}$;
\State $Q_{j}:= [Q_{j-1}, q]$;
\State $r := Aq - \beta-{j}- v_{1}$;
\State $ \alpha_{j}:= q^{*}r $;
\State $r := r - \alpha_{j}q$;
\State $\beta_{j} := krk$;
\If {$(\beta_{j}= 0)$}
   \State $(Q \in F^{n \times j}; \alpha_{1}, . . . ,         \alpha_{j};\beta_{1}, . . . , \beta_{j-1})$
   \EndIf
\EndFor
\end{algorithmic}
\end{algorithm}

\end{document}